# On-Chip Implementation of Cascaded Integrated Comb filters (CIC) for DSP applications

Rozita Teymourzadeh & Prof. Dr. Masuri Othman
VLSI Design Centre
BlokInovasi2, Fakulti Kejuruteraan,
University Kebangsaan Malaysia
43600 UKM, Bangi, Selangor DE, Malaysia
E-mail: **rozita60@vlsi.eng.ukm.my**

*Abstract* - This paper presents the design of a CIC filters based on a low-pass filter for reducing the sampling rate, also known as decimation process. The targeted application for the filter is in the analog to digital conversion (ADC).The CIC is chosen because of its attractive property of both low power and complexity since it dose not required multipliers. Simulink toolbox available in Matlab software is used to design and simulate the functionality of the CIC filter. This paper also shows how sample frequency is decreased by CIC filter and it can be used to give enough stop-band attenuation to prevent aliasing after decimation.

## I. INTRODUCTION

Sigma delta ($\sum\Delta$) modulator is an over sampled modulation technique which provides high resolution sample output in contrast to the standard Nyquist sampling technique. However at the output, the sampling process is needed in order to bring down the high sampling frequency. The CIC filter is a preferred technique for this purpose. Additionally the CIC filter does not require storage for filter coefficients and multipliers as all coefficients are unity [1]. Furthermore its on-chip implementation is efficient because of its regular structure consisting of two basic building blocks, minimum external control and less complicated local timing is required and its change factors is reconfigurable with the addition of a scaling circuit and minimal changes to the filter timing. The CIC filter also removes quantization noise and prevents aliasing introduced during sampling rate decreasing. In this project, the design and implementation of CIC filters, its impulse responses are evaluated by using Matlab. Once the functionality of the filter is verified, the Verilog coding for the filter will be written, simulated and compared with the results from the MATLAB simulations. The chip implementation of the filter will follow. Section II presents overview of the CIC filter followed by section III which gives some results of the characteristics of the filters. Chip implementation of the filters is described in the last section.

## II. An overview of CIC filters

### A. CIC filters to perform the decimation process.

The purpose of the CIC filter is twofold; firstly to remove filtering noise which could be aliased back to the base band signals and secondly to convert high sample rate 1-bit data stream at the output of the Sigma-delta modulator an n-bit data stream with lower sample rate (where n is determined by the specifications of the filters). This process is also known as decimation which is essentially performing the averaging and a rate reduction functions simultaneously. Figure 1 shows the decimation process using the CIC filter (Block in the middle).

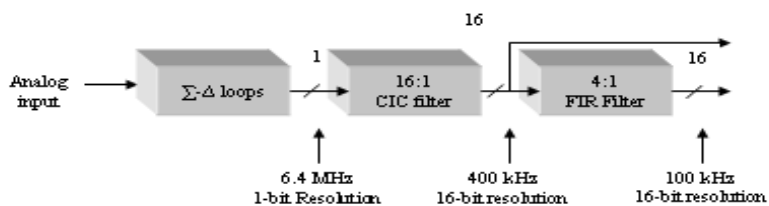

Fig.1: Digital Decimation Process

This CIC filter chosen is a low pass filter with symmetric coefficient value and linear-phase response. It provides 4:1 decimation (in above example) and magnitude compensation for the magnitude change (droop) from the CIC filter output. Detailed analysis will be presented in in section IV.

*B. CIC filters equation and block diagram*

The building blocks of the CIC filter are the integrator (Fig. 2) and the comb filter (Fig. 3).

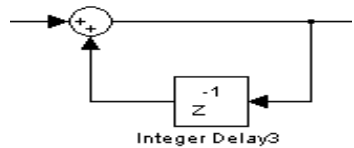

Fig.2: Integrator Cell

For the integrator, its equation in the discrete-time domain and z-domain are:

$$y[n] = y[n-1] + x[n] \qquad 1(a)$$

$$H_I(z) = \frac{1}{1 - z^{-1}} \qquad 1(b)$$

Which is a simply single-pole IIR filter with a unity feedback coefficient and in other words a single integrator by itself is unstable.

The comb filters (another part of CIC filter) which does not require feedback for M (differential delay) = 1 is shown below:

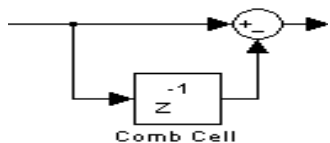

Fig.3: Comb Cell

Sampling rate reduction by the comb filter is given by fs/R where Fs is the high sampling rate and R is the decimation factor of the filter, the operation of the comb filter is:

$$y[n] = x[n] - x[n - RM] \qquad 2(a)$$

$$H_C(z) = (1-z^{-RM}) \qquad (2b)$$

The transfer function after adding Comb and Integrator block and in N stage is given as:

$$H(z) = H_I^N(z) \cdot H_C^N(z) = \frac{(1-z^{-RM})^N}{(1-z^{-1})^N} = \left(\sum_{k=0}^{RM-1} z^{-k}\right)^N \qquad (3)$$

The block diagram of one stage CIC filter is drawn as below:

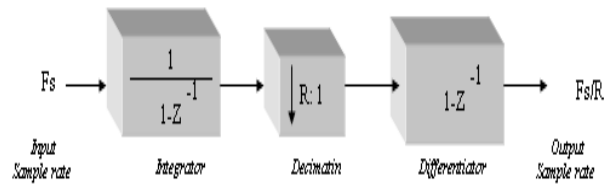

Fig.4: One-stage CIC filtering Process

The gain of the CIC filter can be expressed as:

$$G = (RM)^N \qquad (4)$$

## III. IMPLEMENTATION

### A. Simulation

The structure of the CIC filter in figure 4 is inputted to Simulink toolbox as shown below:

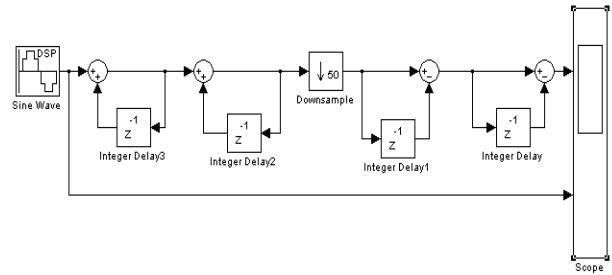

Fig.5: CIC filter simulation

Input and output of simulations (fig.6) make clearer the CIC filter tasks. Input frequency is 200 kHz and output frequency is 4 kHz which give decimation factor of 50. Because of the Integrator block in CIC filter DC gain also has increased and it evaluated by Gain equation: $G = (50 \times 1)^2 = 2500$

So output signal should be have 2500 v DC gain.

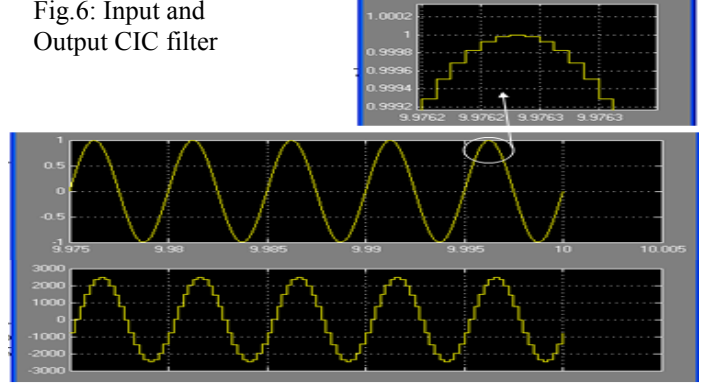

Fig.6: Input and Output CIC filter

## B. Impulse and Frequency Response

Overall filter transfer function of N-order CIC with gain involving can be expressed as :

$$H(z) = \frac{1}{D^N} \frac{(1-z^{-D})^N}{(1-z^{-1})^N} = \left(\sum_{K=0}^{D-1} z^{-k}\right)^N \qquad (5)$$

Where the term $1/D^N$ takes in to account the DC gain of the integrators and $D = RM$ is desired decimation factor.

CIC filters have a linear-phase low-pass frequency response that can be obtained by setting $z = e^{j\omega}$ (by assuming a sampling interval $T = 1/f_s$, the digital frequency $f_d$ is related to ω through ω = $2\pi f T = 2\pi f_d$, where $f$ is the analog frequency, obtaining

$$H(e^{j2\pi f_d}) = \frac{1}{D^N} \left(\frac{Sin(\pi D f_d)}{Sin(\pi f_d)}\right)^N e^{-j2\pi d_d N \frac{D-1}{2}} \qquad (6)$$

The transfer function of equation (6) is equal to zero at integer multiples of the frequency $f_k = k.(1/D)$, with $k = 1,…,[D/2]$ if $D$ is even and $k = 1,…,[D-1/2]$ for $D$ odd.[2]

Impulse response and transfer function detect some of the characteristic of CIC filter for example CIC filter in fig. (7) is stable because its impulse response h(n) decays to 0 as n goes to infinity.

Impulse response has been shown in fig. (7) by using Matlab soft ware:

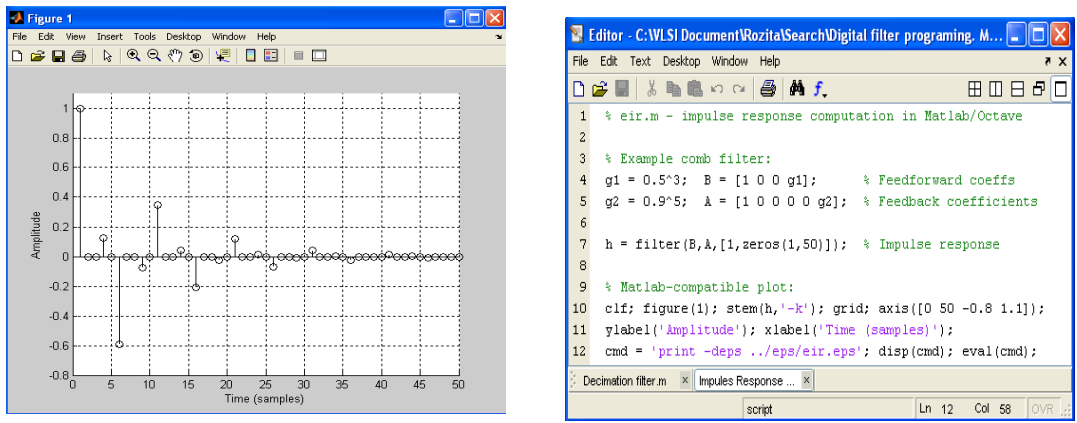

Fig.7: Impulse response

Unstable filters are not normally useful in practice because their output grows exponentially and its energy is not limited.

Fig(8) detects how CIC filter decreases 160 samples to 20 samples by decimation factor equal 8 and also frequency response for a mention filter with M=4, N=4 & R=8 are shown as below:

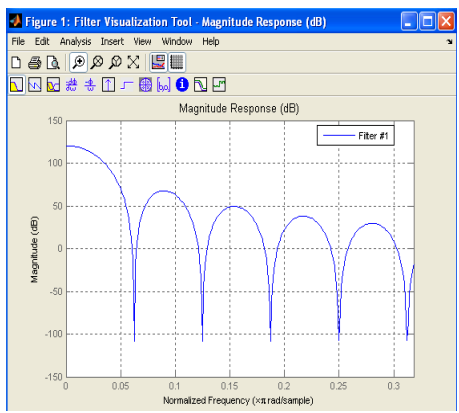
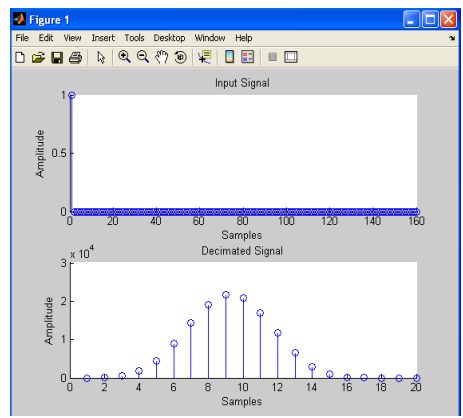

Fig.8: Matlab Programming for frequency and Impulse response

## IV. SPECIFICATION & PASS BAND DROOP

*A. Specification*

CIC filter is equivalent to NFIR filter. Specification of CIC filter has been chosen as table (1):

| Application | |
|---|---|
| Function | amount |
| N – Number of stages | 2 |
| R – Decimation factor | 50 |
| M – Differential Delay | 1 |
| $F_s$ - sample frequency | 200 kHz |
| Fc - band width | 4 kHz |
| $F_{AI}$- Aliasing/imaging attenuation | 20.9 dB |
| Pass band attenuation at Fc (dB) | 1.82 dB |

Tbl.1: Specification of CIC filter

The number of stage in the filter determines the pass band attenuation, Increasing N improves filter ability to reject aliasing but it also increases droop (or roll off) in the filter pass band.[3] Differential delay is used as a design parameter to control the placement of the nulls. For CIC decimation filters the region around every Mth null is folded in to the pass band causing aliasing errors. Differential delay is a filter design parameter used to control the frequency response of filter is restricted to be either 1 or 2. [4]

B. Pass band droop in CIC filter

CIC filters are multiplier less structure, consisting of only adders, sub tractors and register. CIC filter increases droop in pass band frequency and also increasing number of stage increases attenuation at frequency in the locality of the zero. The increases droop may not be acceptable in application. droop is frequently corrected using an additional (non-CIC-based) stage of filtering after the CIC decimator. [4] Because of the pass band droop and therefore narrow usable pass band in CIC filter, many CIC designs utilize an additional FIR filter at the low sampling rate. Filter sharpening can be used to improve the response of CIC filter. This filter will equalize pass band droop and perform a low rate change usually by a factor of two to eight. (CIC filter + Compensation filter= composite filter)

Fig.9: Pass band compensation by FIR filter

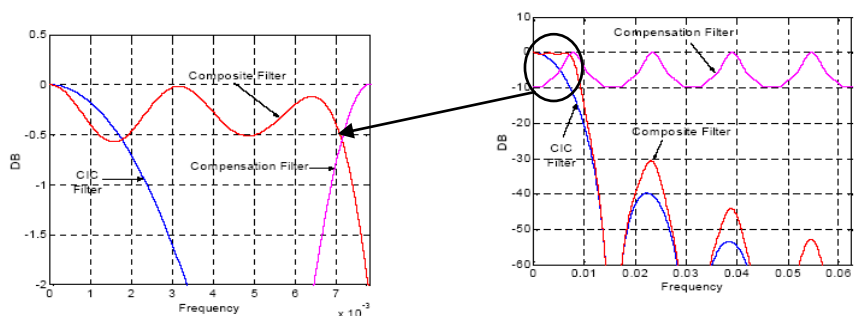

FIR filter or a compensation filter (not part of CIC core) can be used to flatten the pass band frequency response. For a CIC decimator compensation filter operates at the decimated sample rate. Using an appropriate FIR filter in series after the CIC decimation filter can compensate for the induced droop.

## V. CIC FILTER CONSTRAINTS & CHIP DESIGN

### A. CIC filter constraints

CIC decimator filter have the following two constraints:

The word lengths of the filter stages must be no increasing. That is the word length of each filter stage. The number of bits of the first filter stage must be greater than or equal to the quantity Bmax. [5]. Bmax is achieved by followed equation:

$$Bmax = [N \log_2^{RM+B-1}]  \quad (7)$$

Where B is the number of bits of the input.

The number of output bit in last comb stage is achieved as below:

$$Bout = [N \log_2^{RM+B}] \quad (8)$$

Bout bits are needed for each integrator and comb stage.

### B. Chip design

For designing a CIC filter as a chip at FPGA, some pins are needed on it. A simple simulation of this chip has been showed in figure (10):

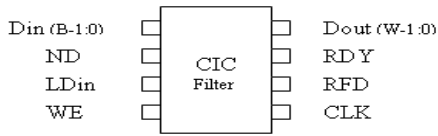

Fig.10: CIC filter Chip

**CLK:** Clock master clock is input pin and become active by rising edge.

**Din:** Data input is input pin and also is B-bit wide filter input port.

**ND:** New Data is Active high and input pin. When this signal is asserted the data sample presented on the Din port is loaded in to the filter.

**Dout:** filter output sample is output pin. W-bit wide filter output sample bus.

**RDY:** filter output sample ready is Active high output pin. Indicates a new filter output sample is available on the Dout port.

**RFD:** Ready For Data is active high output pin. Indicates when the filter can accept a new input sample.

**LDin:** LDin input bus is input pin and used to supply the sample rate change value when the programmable rate change option for the CIC decimator is selected.

**WE:** Write Enable signal is active high input pin. This signal is associated with the LDin port..

This example of CIC filter chip is a basic example and the structure of its action has been considered as chart which has been shown as below:

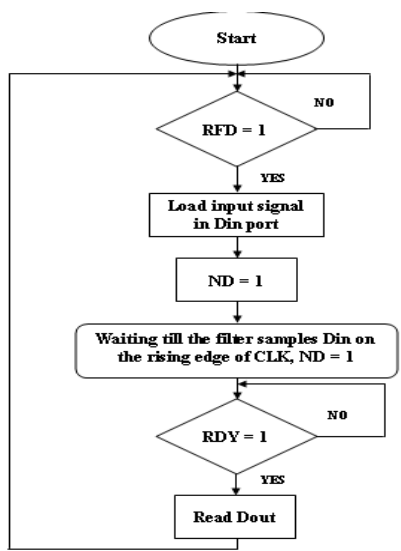

Fig.11: CIC filter Chip

Mention chart shows how pins of chip act together till Dout become ready in output port.

CIC decimation timing has been showed in below figure. In this figure new data sample has been considered to supply to the chip on every clock cycle.

Dout also become ready after passing some clock pulse interval depends on filter latency and R and after that next Dout become ready after R clock cycle.

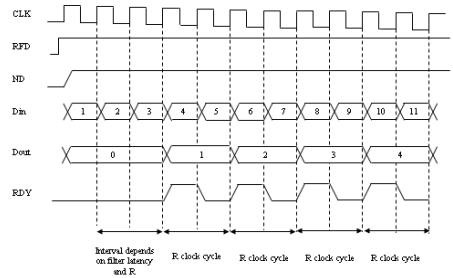

Fig.12: CIC decimation timing

The time interval is a function of the down-sampling factor R and a fixed latency that is related to internal pipeline registers in the CIC chip. The number of pipeline stages depends on the chip customization parameters.[4]

## VI. CONCLUSION

In this paper structure, block diagram and specification of CIC filter by using Matlab software has been proposed and also first step of chip implementation was described. Although the objective of this paper has been written but further work about truncation or rounding which may be used for each stage still has not be done by this paper.

## VII. ACKNOWLEDGEMENT

I wish to thank my supervisor, Prof. Dr. Masuri Othman who has assisted me for completion of this dissertation.

## VIII. REFERENCES

[1] Hogenauer EB, (1981).*An economical class of digital filters for decimation and interpolation,* IEEE transactions on acoustic, speech and signal processing, Sunnyvale, CA.Assp-29(2):155162

[2] Fred Daneshgaran & Massimiliano Laddomada, (2002). *A novel class of decimation filters for sigma delta A/D converters,* Wirel. Commun. Mob. Comput, 867-882.

[3] Matlab Software, (2004). *Mfilt.CICdecim*, filter design toolbox.

[4] Xilinx, (2002).*Logic Core Cascaded Integrator-Comb (CIC) filter V3.0,* Product specification.

[5] Matlab Software, (2004).*CIC decimation.* Signal processing Blockset.